\documentclass[aps, preprint, showkeys]{revtex4}
\usepackage{amsmath}
\usepackage{amsfonts}
\usepackage{graphicx}
\usepackage{amssymb}
\usepackage{float}
\usepackage{graphics}
\usepackage[font=scriptsize]{caption}

\begin{document}

\title{Correlation Dynamics of Qubit-Qutrit Systems in a Classical Dephasing Environment}
\author{G. Karpat}
\email{gkarpat@sabanciuniv.edu}
\affiliation{Faculty of Engineering and Natural Sciences, Sabanci University, Tuzla, Istanbul 34956, Turkey}
\author{Z. Gedik}
\affiliation{Faculty of Engineering and Natural Sciences, Sabanci University, Tuzla, Istanbul 34956, Turkey}

\date{\today}

\begin{abstract}
We study the time evolution of classical and quantum correlations for hybrid qubit-qutrit systems in independent and common classical dephasing environments. Our discussion involves a comparative analysis of the Markovian dynamics of negativity, quantum discord, geometric measure of quantum discord and classical correlation. For the case of independent environments, we have demonstrated the phenomenon of sudden transition between classical and quantum decoherence for qubit-qutrit states. In the common environment case, we have shown that dynamics of quantum and geometric discords might be completely independent of each other for a certain time interval, although they tend to be eventually in accord.
\end{abstract}

\maketitle

\section{Introduction}
Entanglement, being considered as the resource of quantum computation, quantum cryptography and quantum information processing, has been a major focus of research in quantum mechanics [1]. However, recent investigations have demonstrated that quantum entanglement is not the only kind of useful correlation present in quantum states. For instance, the deterministic quantum computation with one qubit does not require entanglement [2]. Moreover, it has been shown both theoretically and experimentally that some separable states might also perform better than their classical counterparts for certain tasks [3-8]. Many different correlation measures have been proposed to detect the nonclassical correlations that cannot be captured by entanglement [9-12]. Among them, quantum discord [9], defined as the difference between quantum versions of two classically equivalent expressions for mutual information, has attracted considerable attention [13-22]. Unfortunately, evaluation of quantum discord requires a potentially complex optimization procedure in general and analytical results have been obtained only in few restricted cases of qubit-qubit and qubit-qudit systems [23-33]. In order to overcome this difficulty, geometric measure of quantum discord has been introduced as an alternative to measure the nearest distance between a given state and the set of zero-discord states [34-37].

Decoherence, the process through which quantum states lose their phase relations due to interactions with the environment, is crucial for understanding the emergence of classicality in quantum systems [38]. One of the most striking features of this unavoidable system-environment interaction on the dynamics of entanglement is the experimentally confirmed [39] phenomenon of the total loss of entanglement in finite time, which is termed as entanglement sudden death (ESD) [39-47]. On the other hand, both Markovian and non-Markovian dynamics of more general quantum and classical correlations have been investigated extensively under various decoherence models [48-76]. Under the conditions where entanglement exhibits a sudden death, quantum discord has been shown to disappear instantaneously in non-Markovian environments [48-50] and  has been observed to resist sudden death in Markovian environments [51]. Another remarkable result first demonstrated in ref. [52], is the existence of a sharp transition between classical and quantum loss of correlations, which has also been experimentally confirmed [53]. This sudden transition implies that there exists a finite time interval, in which only classical correlation is lost and quantum discord is unaffected by noisy environment. Consequently, it has been suggested that quantum discord may be more robust than entanglement, and quantum computation models based on quantum discord correlations might be more relevant than those based on entanglement.

In this work, we consider two different one-parameter families of qubit-qutrit states, interacting with a classical dephasing environment. We study the time evolution of correlations as quantified by negativity, quantum discord, geometric discord and classical correlation both for the cases of multilocal and global dephasing noise. By making a comparative analysis of these correlation measures, we discuss the effects of initial conditions on the dynamics of the qubit-qutrit system. Even for simple one-parameter classes of states, we identify several fundamentally different types of dynamics, including the phenomenon of sudden transition between classical and quantum decoherence.

The outline of this Letter is as follows. In Section 2, we introduce the decoherence model and its solution both for the cases of multilocal and collective dephasing noise. In Section 3, we give a review of the correlation measures used in our discussion. In Section 4, we analyze the dynamics of these correlation measures for certain qubit-qutrit state families. Section 5 includes the summary of our results.

\section{Classical Dephasing Noise}
We consider a composite system of uncoupled spin-$1/2$ and spin-$1$ objects, both of which are under the effect of stochastic environmental fluctuations. The model Hamiltonian we use can be thought as the representative of the class of interactions generating a pure dephasing process [77,78] that is defined as
\begin{eqnarray}
H(t)= -\frac{1}{2} \mu [n_{A}(t)\sigma_{z}^{A}+ n_{B}(t)c_{z}^{B}+n_{AB}(t)(\sigma_{z}^{A}+c_{z}^{B})],
\end{eqnarray}
where we take $\hbar=1$. While $\sigma_{z}$ is the usual Pauli spin operator in z-direction, $c_{z}$ corresponds to z-component of the three level spin
\begin{eqnarray}
c_{z} =
\left(\begin{array}{ccc}
1 & 0  & 0\\
0 & 0  & 0\\
0 & 0  & -1\\
\end{array}\right).
\end{eqnarray}
Here $\mu$ is the gyromagnetic ratio. $n_{i}(t)$ $(i=A, B, AB)$ are stochastic noise fields that lead to statistically independent Markov processes satisfying
\begin{eqnarray}
\langle n_{i}(t) \rangle = 0,  \nonumber
\end{eqnarray}
\begin{eqnarray}
\langle n_{i}(t)n_{i}(t') \rangle = \frac{\Gamma_{i}}{\mu^{2}} \delta(t-t'),
\end{eqnarray}
where $ \langle \cdots \rangle $ stands for ensemble average, and $\Gamma_{i}$ is the damping rate associated with the stochastic field $n_{i}(t)$. The time evolution of the density matrix of the system is given by
\begin{eqnarray}
\rho(t)= \langle U(t) \rho(0) U^{\dagger}(t) \rangle,
\end{eqnarray}
where ensemble averages are evaluated over the three noise fields and the time evolution operator, $U(t)$, is obtained as
\begin{eqnarray}
U(t)= \exp \left[-i \int_0^t \! dt' H(t') \ \right].
\end{eqnarray}
We assume that all the damping parameters are the same $(\Gamma_{A}=\Gamma_{B}=\Gamma_{AB}=\Gamma)$ for the sake of simplicity. First, we focus our attention to the case of multilocal dephasing, i.e., $n_{AB}(t)=0$. In this setting, qubit and qutrit are only interacting with their own environments locally. The resulting time-evolved density matrix in the product basis $\lbrace \vert ij\rangle : i=0, 1, j= 0, 1, 2 \rbrace$ can be written as
\begin{eqnarray}
\rho(t)=
\left(\begin{array}{cccccc}
\rho_{11} & \rho_{12} \gamma & \rho_{13} \gamma^{4} & \rho_{14} \gamma^{4} & \rho_{15} \gamma^{5} & \rho_{16} \gamma^{8}\\
\rho_{21} \gamma & \rho_{22} & \rho_{23} \gamma & \rho_{24} \gamma^{5} & \rho_{25} \gamma^{4} & \rho_{26} \gamma^{5}\\
\rho_{31} \gamma^{4} & \rho_{32} \gamma & \rho_{33} & \rho_{34} \gamma^{8} & \rho_{35} \gamma^{5} & \rho_{36} \gamma^{4}\\
\rho_{41} \gamma^{4} & \rho_{42} \gamma^{5} & \rho_{43} \gamma^{8} & \rho_{44} & \rho_{45} \gamma & \rho_{46} \gamma^{4}\\
\rho_{51} \gamma^{5} & \rho_{52} \gamma^{4} & \rho_{53} \gamma^{5}& \rho_{54} \gamma & \rho_{55} & \rho_{56} \gamma\\
\rho_{61} \gamma^{8} & \rho_{62} \gamma^{5} & \rho_{63} \gamma^{4} & \rho_{64} \gamma^{4} & \rho_{65} \gamma & \rho_{66}\\
\end{array}\right),
\end{eqnarray}
where $ \rho_{ij} $ stands for the elements of the initial density matrix $\rho(0)$ and $\gamma(t)= e^{ -t \Gamma /8}$.
Second, we consider a global dephasing scenario where the spins are interacting with a shared environment collectively and local baths are absent, i.e., $n_{A}(t)=n_{B}(t)=0$. In this case, dynamics of the initial density matrix can be expressed in the same basis as
\begin{eqnarray}
\rho(t)=
\left(\begin{array}{cccccc}
\rho_{11} & \rho_{12} \gamma & \rho_{13} \gamma^{4} & \rho_{14} \gamma^{4} & \rho_{15} \gamma^{9} & \rho_{16} \gamma^{16}\\
\rho_{21} \gamma & \rho_{22} & \rho_{23} \gamma & \rho_{24} \gamma & \rho_{25} \gamma^{4} & \rho_{26} \gamma^{9}\\
\rho_{31} \gamma^{4} & \rho_{32} \gamma & \rho_{33} & \rho_{34} & \rho_{35} \gamma & \rho_{36} \gamma^{4}\\
\rho_{41} \gamma^{4} & \rho_{42} \gamma & \rho_{43} & \rho_{44} & \rho_{45} \gamma & \rho_{46} \gamma^{4}\\
\rho_{51} \gamma^{9} & \rho_{52} \gamma^{4} & \rho_{53} \gamma & \rho_{54} \gamma & \rho_{55} & \rho_{56} \gamma\\
\rho_{61} \gamma^{16} & \rho_{62} \gamma^{9} & \rho_{63} \gamma^{4} & \rho_{64} \gamma^{4} & \rho_{65} \gamma & \rho_{66}\\
\end{array}\right).
\end{eqnarray}
We note that some elements of the initial density matrix $\rho(0)$ are not affected by decoherence in collective dephasing setting, which is an indicator of the existence of decoherence-free subspaces.

\section{Measures of Correlations}
Before starting to discuss the dynamics of quantum correlations, we review the correlation measures used in our investigation, namely, negativity, quantum discord and geometric measure of quantum discord.
The quantification of entanglement is well understood for the case of two-qubits [79,80], but little is known about its generalization to higher dimensional bipartite mixed states. Negativity is an entanglement measure which can be easily calculated for any bipartite entangled state having negative partial transpose in all dimensions. Although we cannot conclude whether a positive partial transpose (PPT) state (zero negativity state) is entangled or separable in general, it has been shown that all PPT states of qubit-qubit and qubit-qutrit systems are separable [81,82]. Thus, qubit-qutrit entanglement is fully characterized by negativity. For a given density matrix $\rho^{AB}$, negativity can be defined as twice the absolute sum of the negative eigenvalues of partial transpose of $\rho^{AB}$ with respect to the smaller dimensional system,
\begin{eqnarray}
N(\rho^{AB})=\sum_{i}|\eta_{i}|-\eta_{i},
\end{eqnarray}
where $\eta_{i}$ are all of the eigenvalues of $(\rho^{AB})^{T_{A}}$. This definition ensures that negativity of maximally entangled qubit-qutrit states are normalized to one.

The total amount of quantum and classical correlations in a qubit-qutrit state can be obtained without difficulty by evaluating the quantum mutual information which is defined as
\begin{eqnarray}
I(\rho^{AB})=S(\rho^{A})+S(\rho^{B})-S(\rho^{AB}),
\end{eqnarray}
where $\rho^{AB}$ and $\rho^{k}$ $(k=A, B)$ are the density matrix of the total system and reduced density matrix of subsystems, respectively, and $S(\rho)=-Tr(\rho \textmd{log}_{2} \rho)$ is the von-Neumann entropy. On the other hand, a measure of classical correlations contained in a quantum state is provided by [9,10]
\begin{eqnarray}
C(\rho^{AB})= S(\rho^{B})-\min_{\{\Pi_{k}^{A}\}}\sum_{k}p_{k}S(\rho_{k}^{B}),
\end{eqnarray}
where $\{\Pi_{k}^{A}\}$ defines a set of orthonormal projectors (a von-Neumann measurement), performed on subsystem $A$ and $\rho_{k}^{B}=Tr_{A}((\Pi_{k}^{A} \otimes I^{B})\rho^{AB})/p_{k}$ is the remaining state of subsystem $B$ after obtaining the outcome $k$ with the probability $p_{k}=Tr((\Pi_{k}^{A} \otimes I^{B})\rho^{AB})$. We intend to evaluate $C(\rho^{AB})$ for qubit-qutrit states assuming that the measurement is performed on the qubit part of the hybrid system. A von-Neumann measurement $\{\Pi_{1}^{A}, \Pi_{2}^{A}\}$ can be represented by
\begin{eqnarray}
\Pi_{1}^{A}=\frac{1}{2}\left(I^{A}_{2}+\sum_{j=1}^{3}n_{j}\sigma_{j}^{A}\right),  \nonumber \\
\Pi_{2}^{A}=\frac{1}{2}\left(I^{A}_{2}-\sum_{j=1}^{3}n_{j}\sigma_{j}^{A}\right),
\end{eqnarray}
where $\sigma_{j}(j=1,2,3)$ are the usual Pauli spin operators and $n=(\sin\theta\cos\phi,\sin\theta\sin\phi,\cos\theta)^{T}$ is a unit vector on the Bloch sphere with $\theta \in [0,\pi)$ and $\phi \in [0,2\pi)$. Quantum discord [9], which measures the amount of quantum correlations, is then defined as the difference between total and classical correlations
\begin{eqnarray}
D(\rho^{AB})=I(\rho^{AB})-C(\rho^{AB}).
\end{eqnarray}
It is possible to show that quantum discord is not a symmetric quantity in general, meaning its value depends on whether the measurement is performed on subsystem A or B. Since the calculation of classical correlation involves a potentially complex optimization process, there exists no general analytical expression of discord even for the simplest case of two-qubit states. For the relatively simple qubit-qutrit mixed states used in our study, we will obtain the quantum discord via numerical optimization of the von-Neumann measurements, which will include a minimization over two independent real parameters $\theta$ and $\phi$.

In order to overcome the difficulties experienced with the analytical calculation of quantum discord, an alternative geometrized version called geometric measure of quantum discord has been proposed [34]. It measures the nearest distance between a given state and the set of zero-discord states. Geometric discord can be mathematically defined as
\begin{eqnarray}
D^{g}(\rho^{AB})=\min_{\chi}\| \rho^{AB}-\chi \|^{2},
\end{eqnarray}
where the minimum is over the set of zero-discord states and the geometric quantity $\|X-Y \|^{2}=Tr(X-Y)^{2}$ denotes the square of the Hilbert-Schmidt norm. A state $\chi$ on $H^{A}\otimes H^{B}$ has vanishing discord if and only if it is a classical-quantum state, that is
\begin{eqnarray}
\chi=\sum_{k=1}^{m}p_{k}|k\rangle\langle k| \otimes \rho_{k}
\end{eqnarray}
where $\{p_{k}\}$ is a probability distribution, $\{|k\rangle\}$ is an arbitrary orthonormal basis for $H^{A}$ and $\rho_{k}$ is a set of arbitrary density operators on $H^{B}$. Recently, an exact analytical formula has been obtained for the geometric discord of a bipartite state of $2 \times n$ dimensions [35,37]. Since our discussion only involves qubit-qutrit states, we focus on the case of $2 \times 3$ states. The density operators acting on a bipartite system $H^{A}\otimes H^{B}$ with $\textmd{dim}H^{A}=2$ and $\textmd{dim}H^{B}=3$ can be represented as
\begin{eqnarray}
\rho^{AB}=\frac{1}{6}\left(I_{6}+\sum_{i=1}^{3}x_{i}\sigma_{i} \otimes I_{3}+\sum_{j=1}^{8}y_{j} I_{2} \otimes \lambda_{j}+\sum_{i=1}^{3}\sum_{j=1}^{8}t_{ij} \sigma_{i} \otimes \lambda_{j}\right)
\end{eqnarray}
where $\sigma_{i}$ and $\lambda_{j}$ are the traceless Hermitian generators of $SU(2)$ and $SU(3)$, respectively, satisfying $Tr(\sigma_{i}\sigma_{j})=Tr(\lambda_{i}\lambda_{j})=2\delta_{ij}$.
The components of the local Bloch vectors $x_{i}, y_{j}$ and the correlation matrix $T$ can be calculated as
\begin{eqnarray}
x_{i}=Tr(\rho^{AB}(\sigma_{i} \otimes I_{3}))  \nonumber \\
y_{j}=\frac{3}{2}Tr(\rho^{AB}(I_{2} \otimes \lambda_{j}))  \nonumber \\
T=t_{ij}=\frac{3}{2}Tr(\rho^{AB}(\sigma_{i} \otimes \lambda_{j}))
\end{eqnarray}
Then, the exact formula for geometric discord of qubit-qutrit states can be written as
\begin{eqnarray}
D^{g}(\rho^{AB})=\frac{1}{6}\|x\|^{2}+\frac{1}{9}\|T\|^{2}-k_{max}
\end{eqnarray}
where $x=(x_{1},x_{2},x_{3})^{T}$ and $k_{max}$ is the greatest eigenvalue of the matrix ($\frac{xx^{T}}{6}+\frac{TT^{T}}{9})$. Since geometric discord is not normalized to one and its value for maximally entangled qubit-qutrit states are $0.5$, we will consider the quantity $2D^{g}(\rho^{AB})$ as we compare the geometric discord with other correlation measures.

\section{Dynamics of Correlations}
In the following sections, we will investigate the correlation dynamics for two different one-parameter families of qubit-qutrit states: entangled $\rho_{e}(p)$ and separable $\rho_{s}(r)$ defined by
\begin{eqnarray}
\rho_{e}(p)=\frac{p}{2}(|00\rangle\langle00|+|01\rangle\langle01|+|00\rangle\langle12|+
|11\rangle\langle11|+|12\rangle\langle12| \nonumber \\
+|12\rangle\langle00|)+\frac{1-2p}{2}(|02\rangle\langle02|+
|02\rangle\langle10|+|10\rangle\langle02|+|10\rangle\langle10|)
\end{eqnarray}
\begin{eqnarray}
\rho_{s}(r)=\frac{r}{2}(|00\rangle\langle00|+|01\rangle\langle01|+|00\rangle\langle12|+
|11\rangle\langle11|+|12\rangle\langle12| \nonumber \\
+|12\rangle\langle00|+|02\rangle\langle10|+|10\rangle\langle02|)
+\frac{1-2r}{2}(|02\rangle\langle02|+|10\rangle\langle10|)
\end{eqnarray}
where $0\leq p\leq1/2$ and $0\leq r\leq1/3$. Note that the entangled family $\rho_{e}(p)$ is separable only for $p=1/3$.

\subsection{Correlations under Multilocal Dephasing}
We first discuss the time evolution of correlations under multilocal classical dephasing noise. The separable family $\rho_{s}(r)$ naturally contains no entanglement since it has PPT for all possible values of $r$. Negativity of the entangled family  $\rho_{e}(p)$ is given as
\begin{eqnarray}
N_{e}(p,\tilde{\gamma})=\frac{1}{2}[|p(1+2\tilde{\gamma})-\tilde{\gamma}|+|p(2+\tilde{\gamma})-1|
-(p-1)(\tilde{\gamma}-1)],
\end{eqnarray}
where $\tilde{\gamma}(t)=e^{ -t \Gamma}$. On the other hand, both of the families have non-vanishing geometric discord in general, which can be calculated as
\begin{eqnarray}\nonumber
D^{g}_{e}(p,\tilde{\gamma}) &=& \frac{1}{4}[1+2\tilde{\gamma}^{2}-2p(3+4\tilde{\gamma}^{2})+p^{2}(9+10\tilde{\gamma}^{2}) \\ &&-\max\{(1-3p)^{2},(1-3p)^{2}\tilde{\gamma}^{2},(1-p)^{2}\tilde{\gamma}^{2}\}],
\end{eqnarray}
\begin{eqnarray}
D^{g}_{s}(r,\tilde{\gamma})=\frac{1}{4}[1 - 6r+r^{2}(9+4\tilde{\gamma}^{2}) -\max\{(1-3r)^{2}, 4r^{2}\tilde{\gamma}^{2}\}].
\end{eqnarray}

\textit{Dynamics of the entangled family.}
We start our investigation by considering $\rho_{e}(0)$ and $\rho_{e}(1/2)$. Correlation dynamics of these two states are completely different from the other members of the family. For $\rho_{e}(1/2)$, Fig. 1(a) displays that while classical correlation is not affected by external noise, all three quantum correlations decay in a monotonic fashion. In this case, negativity seems to be more robust than quantum and geometric discords. On the other hand, $\rho_{e}(0)$ is a maximally entangled state and its general behavior is almost the same as $\rho_{e}(1/2)$ except all of its correlations are one initially. Dynamics of the correlations for the remaining members of the family are far more interesting. For all of the states corresponding to the regime $1/2>p>0$ (excluding $p=1/3$), entanglement disappears in a finite time suffering ESD. More important, we observe the sudden transition from classical to quantum decoherence [52], i.e, there exists a critical instant $t_{c}$ at which the quantum state stops losing classical correlation and starts losing quantum discord. Geometric discord fails to keep up with quantum discord in the classical decoherence region, but its decay still suddenly hastens at the critical time $t_{c}$. Fig. 1(b) shows an example of this behavior for $p=0.25$. By choosing different initial states from this family, it is possible to prolong the time interval in which quantum discord remains constant but there exists a trade-off between the initial magnitude of the quantum discord and its survival time. For instance, Fig. 1(c) illustrates the case for $p=0.2$. Although sudden changes of all correlation measures occur at the same time instant for all the initial states considered in our study, this is not a general feature of all quantum states. Examples of states have been presented in ref. [73] for which evolutions of quantum and geometric discords are not affected by the discontinuities in each others dynamics.

\textit{Dynamics of the separable family.}
The two end points of this family, namely, $\rho_{s}(0)$ and $\rho_{s}(1/3)$, are not particulary interesting since they do not contain any kind of quantum correlations. For the initial states corresponding to the interval $1/5\geq r>0$, classical correlation does not feel the noise fields, whereas quantum and geometric discords decay monotonically. However, the regime $1/3>r>1/5$ is definitely more interesting, since we observe an analogue of the sudden transition from classical to quantum decoherence. In this case, though quantum discord is not constant and decays together with the classical correlation, we notice that geometric discord is unaffected by environment for a finite time interval. In other words, there exists an instant of time $\tilde{t}_{c}$ at which the system stops losing classical correlation and starts losing geometric discord. An example is presented in Fig. 1(d), where $r=0.25$ and the critical time $\tilde{t}_{c}=\ln2/\Gamma$. Note that the state keeps losing quantum discord throughout the dynamics but as soon as $\tilde{t}_{c}$ is reached, the decay rate of quantum discord hastens.
\begin{figure}[H]
\begin{center}
\includegraphics[scale=0.8]{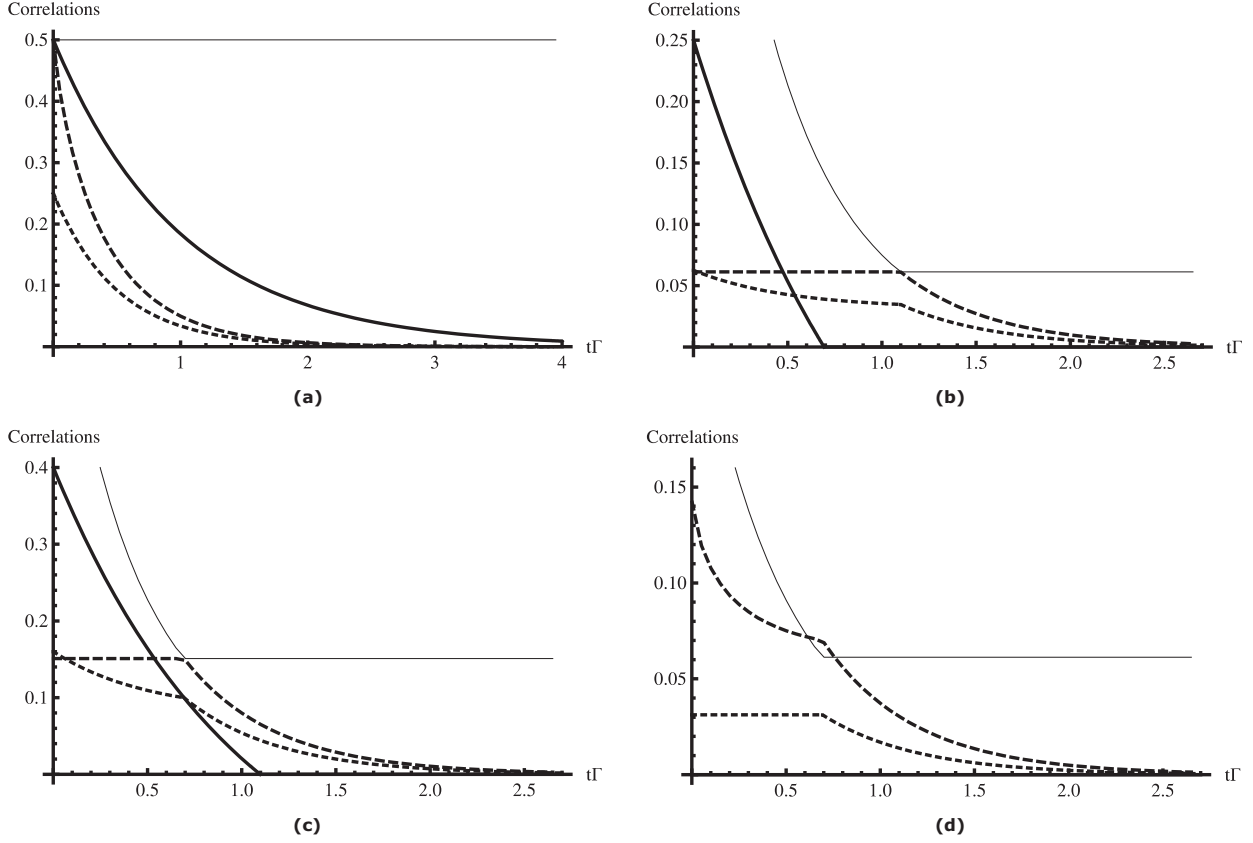}
\caption{Dynamics of negativity $N$ (thick solid line), geometric discord $2D^{g}$ (dotted line), numerically evaluated quantum discord $D$ (dashed line) and classical correlation $C$ (thin solid line) as a function of the dimensionless parameter $t\Gamma$ under the effect of multilocal classical dephasing noise. The initial states are $\rho_{e}(p)$ with (a) $p=0.5$ (b) $p=0.25$ (c) $p=0.2$ and $\rho_{s}(r)$ for (d) $r=0.25$.}
\end{center}
\end{figure}

\subsection{Correlations under Collective Dephasing}
In this section, we discuss the time evolution of correlations under collective classical dephasing noise. Negativity of the entangled family $\rho_{e}(p)$ reads as
\begin{eqnarray}
N_{e}(p,\tilde{\gamma})= \frac{1}{2}[|3p-1|+|p(2+\tilde{\gamma}^{2})-1|
-p(1-\tilde{\gamma}^{2})].
\end{eqnarray}
Geometric discord for the two families are also obtained as
\begin{eqnarray}\nonumber
D^{g}_{e}(p,\tilde{\gamma}) &=& \frac{1}{4}[3-14p+p^{2}(17+2\tilde{\gamma}^{4}) -\max\{(1-3p)^{2}, \\
&& (p(\tilde{\gamma}^{2}-2)+1)^{2},(p(\tilde{\gamma}^{2}+2)-1)^{2}\}],
\end{eqnarray}
\begin{eqnarray}\nonumber
D^{g}_{s}(r,\tilde{\gamma}) &=& \frac{1}{4}[1 - 6r+r^{2}(11+2\tilde{\gamma}^{4}) -\max\{(1-3r)^{2}, \\ && r^{2}(1-\tilde{\gamma}^{2})^{2},r^{2}(1+\tilde{\gamma}^{2})^{2}\}].
\end{eqnarray}

\textit{Dynamics of the entangled family.}
The correlation dynamics of the entangled family under collective noise is a lot richer than its dynamics under multilocal noise. While all of the correlations hold unchanged for the maximally entangled state $\rho_{e}(0)$, correlation dynamics of the state $\rho_{e}(1/2)$ is no different than what's described in Fig. 1(a) except for the fact that correlations decay faster. In the regime $1/3\geq p>0$, quantum and geometric discords are both uniformly amplified and become stable after a certain point. Negativity is conserved since this regime consists of disentanglement-free states. Fig. 2(a) displays an example of this case for $p=0.2$. Classical correlation, which can be greater or smaller than quantum discord, decreases monotonically and gets stable as well. For the regime $2/5\geq p>1/3$, behaviors of classical correlation and quantum discord aren't changed. On the other hand, geometric discord acquires a minimum without a sudden change. Although all other correlations survive the effects of the environment, negativity disappears in a finite time suffering sudden death. Fig. 2(b) illustrates the situation for $p=0.4$. It is noteworthy that geometric discord can decrease as quantum discord increases. Next, we examine the interval $1/2>p>2/5$. Whereas the states keep experiencing ESD, all other correlations show sudden changes in their evolutions at the same instant. Fig. 2(c) gives an example this behavior for $p=0.45$. Note that geometric discord and classical correlation diminish as quantum discord gets amplified until a critical time is reached. After that instant, both quantum and geometric discords start to weaken until they reach a stable value, but classical correlation is not affected by noise at all.

\textit{Dynamics of the separable family.}
Starting with $\rho_{s}(1/3)$, we immediately see that smooth amplification of both quantum and geometric discords is possible in this setting. In the regime $1/5\geq r>0$, classical correlation is unaffected by noise but quantum and geometric discords decay in a monotonic way until they eventually become stable. In the interval $1/4>r>1/5$, all correlations start to evolve in a different fashion but they all become discontinuous simultaneously at a certain critical instant. After that instant, classical correlation becomes constant as other measures starts to decrease until they finally get stable. Fig. 2(d) illustrates this behavior for $r=0.23$.
\begin{figure}[H]
\begin{center}
\includegraphics[scale=0.8]{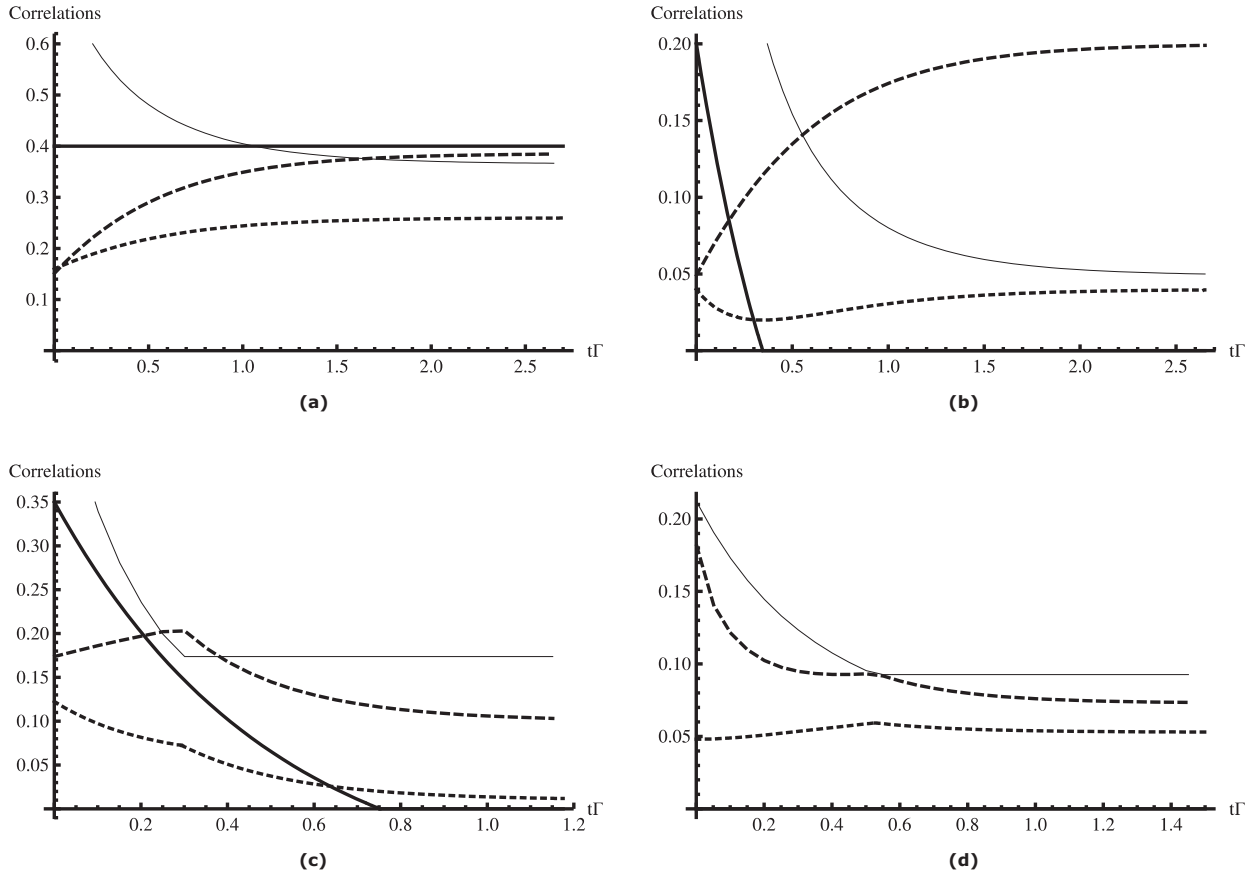}
\caption{Dynamics of negativity $N$ (thick solid line), geometric discord $2D^{g}$ (dotted line), numerically evaluated quantum discord $D$ (dashed line) and classical correlation $C$ (thin solid line) as a function of the dimensionless parameter $t\Gamma$ under the effect of collective classical dephasing noise. The initial states are $\rho_{e}(p)$ with (a) $p=0.2$ (b) $p=0.4$ (c) $p=0.45$ and $\rho_{s}(r)$ for (d) $r=0.23$.}
\end{center}
\end{figure}

\section{Conclusion}
In summary, we have analyzed the time evolution of negativity, quantum discord, geometric discord and classical correlation for two different one-parameter families of qubit-qutrit states, assuming that the states are in a classical dephasing environment. Considering the cases of multilocal and collective dephasing separately, we have noticed that dynamics of correlations are strongly dependent on the initial conditions even for one-parameter families of states. Under multilocal noise, we have demonstrated the phenomenon of sudden transition between classical and quantum decoherence for hybrid qubit-qutrit systems extending the results of ref. [52]. In fact, this transition might be a generic feature existing in all bipartite quantum systems but a definitive demonstration would require an analytic expression for quantum discord in arbitrary dimensions. Furthermore, for a class of separable states, we have observed an analogue of this phenomenon. Namely, geometric discord remains constant for a finite time interval while classical correlation decreases, then when geometric discord starts to decay, classical correlation becomes constant. Under global noise, dynamics of correlations are quite diverse. We have shown that although quantum and geometric discords can evolve initially completely independent of each other for a certain time period, they tend to be eventually in accord. Smooth amplification of quantum and geometric discords is also possible in this case. On the other hand, we have confirmed that entanglement as quantified by negativity can suffer sudden death for qubit-qutrit states both in global and multilocal dephasing settings. Our findings clearly indicate that different measures of quantum correlations are conceptually different. Lastly, it should be possible to prolong the survival time of correlations by considering non-Markovian extensions such as Ornstein-Uhlenbeck type of noise models [65].

\section*{Acknowledgements}
This work has been partially supported by the Scientific and Technological Research Council of Turkey (TUBITAK) under Grant 111T232. The authors would like to thank B. \c{C}akmak for helpful discussions.

\end{document}